\documentclass[aps,pra,showpacs,amsmath,amssymb,amsfonts,lengthcheck,twocolumn,longbibliography,superscriptaddress]{revtex4-2}

\usepackage{graphicx}
\usepackage{subcaption}
\usepackage{caption}
\usepackage{dsfont}

\usepackage{verbatim}
\usepackage{dcolumn}
\usepackage{bm}
\usepackage{epsf}
\usepackage{color}
\usepackage[toc]{appendix}
\usepackage{stmaryrd}
\usepackage{amsthm}
\usepackage{hhline}
\usepackage{amssymb}
\usepackage{mwe,tikz}\usepackage[percent]{overpic}
\usepackage{tikz}
\usetikzlibrary{arrows,shapes,calc,matrix}
\usepackage[normalem]{ulem}
\expandafter\let\csname equation*\endcsname\relax
\expandafter\let\csname endequation*\endcsname\relax
\usepackage{amsmath}
\usepackage{xstring}

\makeatletter
\newcommand\footnoteref[1]{\protected@xdef\@thefnmark{\ref{#1}}\@footnotemark}
\makeatother

\newcommand\hocom[1]{}

\newcommand{\ba}{\begin{eqnarray}}
	\newcommand{\ea}{\end{eqnarray}}
\newcommand{\bmath}{\begin{mathletters}}
	\newcommand{\emath}{\end{mathletters}}
\newcommand{\ban}{\begin{eqnarray*}}
	\newcommand{\ean}{\end{eqnarray*}}

\newcommand{\bra}[1]{\left\langle #1\right|}
\newcommand{\ket}[1]{\left|#1\right\rangle}
\newcommand{\braket}[2]{\left\langle #1|#2\right\rangle}
\newcommand{\ketbrad}[1]{|#1\rangle\!\langle #1|}
\newcommand{\tr}[1]{\mathrm{tr}\left\{#1\right\}}
\newcommand{\ptr}[2]{\mathrm{tr_{#1}}\left\{#2\right\}}

\newcommand{\la}{\left\langle}
\newcommand{\ra}{\right\rangle}

\newcommand{\e}[1]{\exp{\left(#1\right)}}

\newcommand{\id}{\mathbb{I}}
\newcommand{\com}[2]{\left[#1,\,#2\right]}

\newcommand{\bla}{bla\\bla\\bla\\bla\\bla}

\newcommand{\mc}[1]{\mathcal{#1}}

\newcommand{\mf}[1]{\mathfrak{#1}}
\newcommand{\mrm}[1]{\mathrm{#1}}

\usepackage{fmtcount}

\newcommand{\draftmode}{1}    
\newcommand{\notetoself}[1]{\ifnum \draftmode=1 {\color[rgb]{0,0,0.8} [#1]} \fi}  
\newcommand{\cuttext}[1]{\ifnum \draftmode=1 {\color[rgb]{0,0.5,0} [#1]} \fi}  
\newcommand{\warntext}[1]{\ifnum \draftmode=1 {\color[rgb]{0.9,0.6,0} #1} \else {#1} \color{black} \fi}
\newcommand{\aref}[1]{{Appendix~\hyperref[#1]{A}}}
\newcommand{\bref}[1]{{Appendix~\hyperref[#1]{B}}}
\newcommand{\dref}[1]{{Appendix~\hyperref[#1]{C}}}
\usepackage[colorlinks=true,citecolor=blue,linkcolor=blue,urlcolor=blue]{hyperref}
\usepackage{cleveref}

\DeclareMathOperator{\arctanh}{arctanh}

\begin{document}

\title{Pointer states and quantum Darwinism with 2-body interactions}
\author{Paul Duruisseau}
 \email{paul.duruisseau@ens-paris-saclay.fr}
\affiliation{
 ENS Paris-Saclay, 91190, Gif-sur-Yvette, France
}

\author{Akram Touil}
 \email{atouil@lanl.gov}
  \affiliation{Theoretical Division, Los Alamos National Laboratory, Los Alamos, New Mexico 87545
}
 \affiliation{Center for Nonlinear Studies, Los Alamos National Laboratory, Los Alamos, New Mexico 87545
}
\author{Sebastian Deffner}
 \email{deffner@umbc.edu}
\affiliation{
 Department of Physics, University of Maryland, Baltimore County, Baltimore, MD 21250, USA
}

\date{\today}

\begin{abstract}
Quantum Darwinism explains the emergence of classical objectivity within a quantum universe. However, to date most research in quantum Darwinism has focused on specific models and their stationary properties. To further our understanding of the quantum-to-classical transition it appears desirable to identify the general criteria a Hamiltonian has to fulfill to support classical reality. To this end, we categorize all models with 2-body interactions, and we show that only those with separable interaction of system and environment can support a pointer basis. We further show that ``perfect'' quantum Darwinism can only emerge if there are no intra-environmental interactions. Our analysis is complemented by the solution of the ensuing dynamics. We find that in systems that exhibit information scrambling, the dynamical emergence of classical objectivity is in direct competition with the non-local spread of quantum correlations. Our rigorous findings are illustrated with the numerical analysis of four representative models.
\end{abstract}

\maketitle

\section{Introduction}

We live in a quantum universe, yet our everyday reality is well-described by classical physics. Hence, the obvious question to ask is where all the quantum information and correlations hide. The quantum nature of our universe is captured by its ability to be in a superposition of classically allowed states. The transition from quantum to classical is a two-step process. The first, necessary but not sufficient, step is the destruction of quantum superpositions, i.e., destruction of all interference phenomena. The theory of decoherence teaches us that it is the interaction between the quantum system and its environment that is the cause of this phenomenon \cite{Schlosshauer2019PR}. The destruction of quantum superpositions presupposes a privileged and unique quantum basis. The elements of this basis are called pointer states \cite{pointerstates1,pointerstates2,pointerstates3}. Any quantum superposition written in this basis decomposes into a classical mixture under the effect of environmental interaction. Thus, pointer states are precisely the only quantum states that are stable under this interaction. 

Quantum Darwinism \cite{Zurek2000AP,PhysRevA.72.042113,PhysRevLett.93.220401,RevModPhys.75.715,zurek_2009,girolami_22,touil2022branching,blume_2006,Deffner2023book} builds on decoherence theory and goes a step further, approaching the problem from the point of view of quantum information theory. An outside observer has no direct access to a system of interest $\mf{S}$, but rather the environment $\mf{E}$ acts as a communication channel. Since any real environment is tremendously large, ``observing''  $\mf{S}$ actually means that an observer intercepts only a small, possibly even tiny fragment $\mf{F}$ of $\mf{E}$, and then reconstructs the state of $\mf{S}$ from the information carried by $\mf{F}$.
 
If the constituents of the environment do not interact, such as, for instance, photons \cite{photon,photon2}, then the information about $\mf{S}$ is accessible by \emph{local} measurements on $\mf{E}$. However, reality is a little more complicated, and in general, the constituents of $\mf{E}$ do interact. Such intra-environmental interactions  lead to the build-up of non-local correlations, which is the root-cause of information scrambling \cite{Hayden2007JHEP,Swingle2018NP,Touil2020QST,Bao2021JHEP,Chen2022RP,Fisher2023}. Thus, an observer has to access a macroscopic fraction of $\mc{E}$ to reconstruct unambiguous information about $\mf{S}$.

Despite the significant attention scrambling dynamics has received in the literature \cite{Hayden2007JHEP,Swingle2018NP,Touil2020QST,Bao2021JHEP,Chen2022RP,Fisher2023,swingle_2016,Yoshia2019PRX}, curiously little is known about the quantum-to-classical transition in the presence of scrambling. Only rather recently, several studies have started to unveil the interplay of decoherence and scrambling \cite{Yoshia2019PRX,Xu2019PRL,Touil2021PRXQ,Zanardi2021PRA,Xu2021PRB,Dominguez2021PRA,Cornelius2022PRL,Han2022Entropy,Andreadakis2023PRA}. 

More directly relevant to our present work is Ref.~\cite{Riedel_2012}, which analyzed a specific model where intra-environmental interactions scramble the information encoded in different fragments $\mf{F}$. This scenario, scrambling only in $\mf{E}$, but  not in $\mf{S}$, makes it easier to highlight the competition between the local transfer of information from $\mf{S}$ to each degree of freedom of $\mf{E}$, and the scrambling of information between the different $\mf{F}$ of $\mf{E}$ due to their interactions.

Current research in quantum Darwinism \cite{Deffner2023book} is driven by the analysis of increasingly complex model systems. However, the focus has remained on particular qubit-models \cite{exmodel1,exmodel2,exmodel3,exmodel4,exmodel5,Riedel_2012}, since their dynamics is tractable. Despite, or rather because of continued progress in our understanding it appears desirable to elucidate the general properties of Hamiltonians that support the emergence of quantum Darwinism. More precisely, it is instrumental to sort all possible interacting many-body Hamiltonians into classes that support a pointer basis for $\mf{S}$, and which sub-classes of these will further exhibit the emergence of classical objectivity. Such a classification will also unveil if and under what circumstances, quantum Darwinism can emerge in the presence of scrambling dynamics.

In the present work, we consider a qubit of interest $\mf{S}$, that interacts with an environment $\mf{E}$ also comprised of qubits. Hence, scrambling of information may only occur in $\mf{E}$, but not in $\mf{S}$. Further, for the sake of simplicity we restrict ourselves to arbitrary two-body interactions.

\begin{figure}
\centering
   \begin{subfigure}[b]{.5\textwidth}
        \includegraphics[width=.75\textwidth]{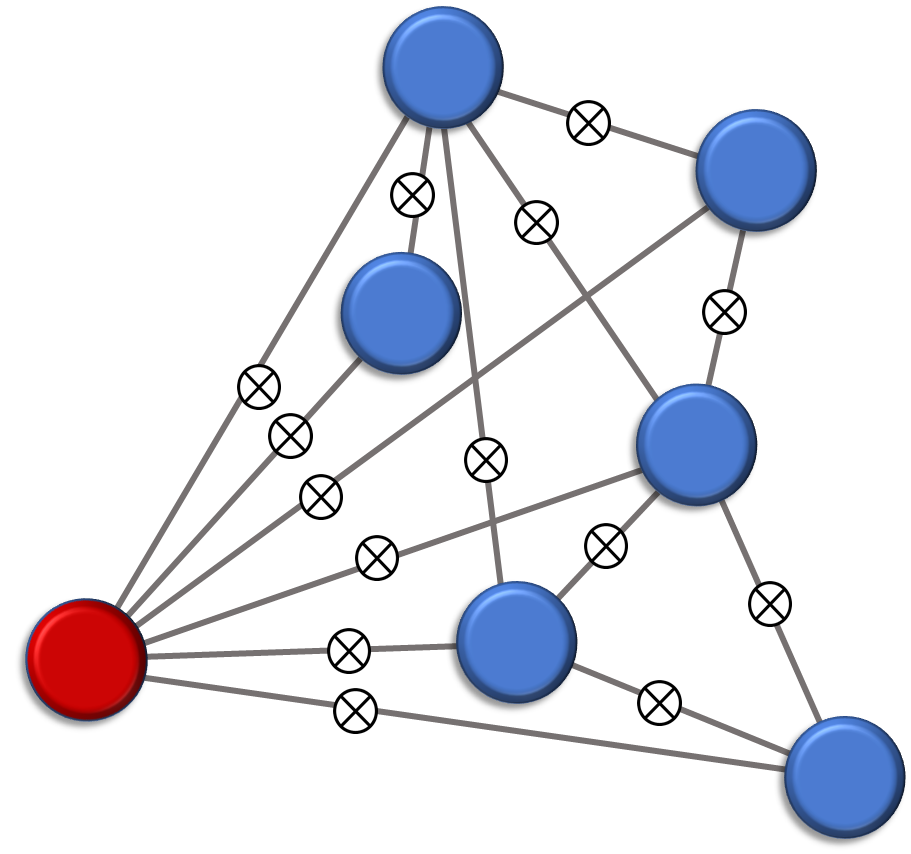}
        \subcaption{Arbitrary two-body interaction model.}
        \label{fig:schema1}
    \end{subfigure}
    \begin{subfigure}[b]{.5\textwidth}
       \includegraphics[width=.75\textwidth]{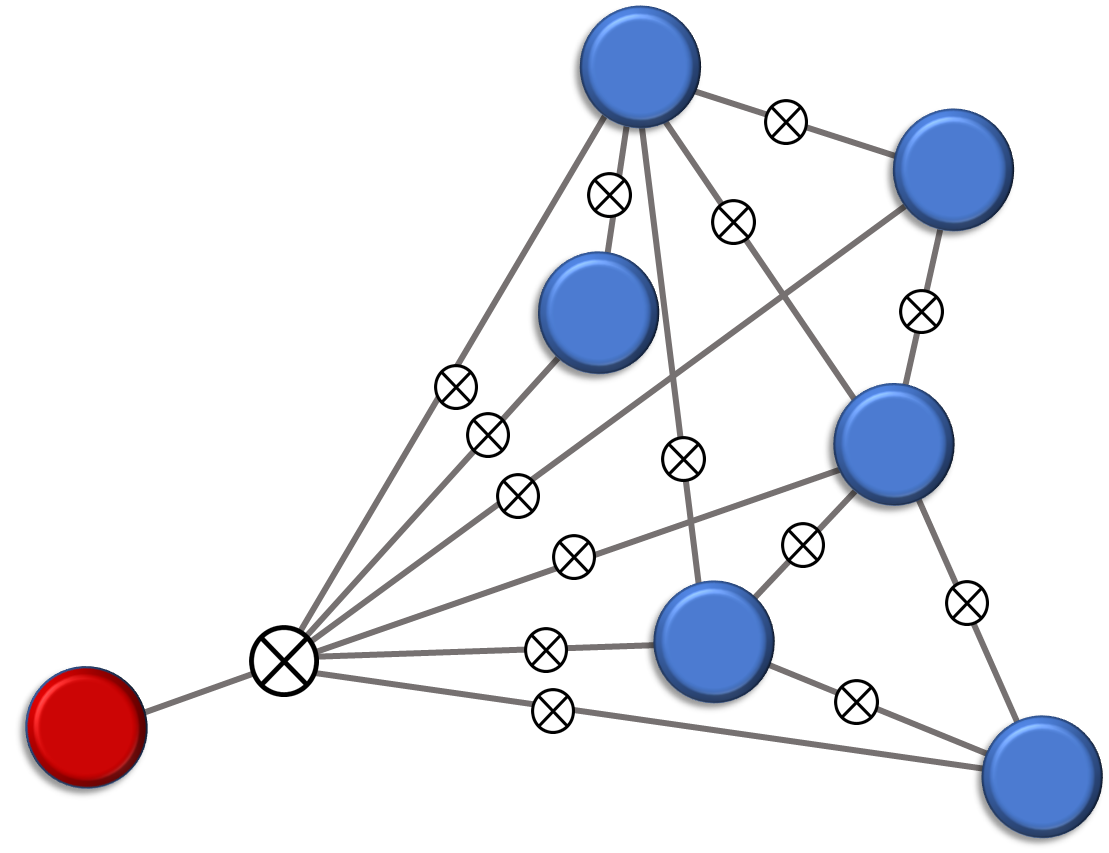}
       \subcaption{$\mf{S}$-$\mf{E}$ separable two-body interaction model.}
       \label{fig:schema2}
    \end{subfigure}
\caption{Two-body interactions between $\mf{S}$ (red) and $\mf{E}$ (blue). Lines depict interaction terms.}
\label{fig:schemas}
\end{figure}

In a first part of our analysis, we show that the existence of a pointer basis for $\mf{S}$ imposes a specific structure for the total Hamiltonian describing the evolution of the universe $\mf{S}\otimes\mf{E}$. In fact, we will see that a pointer basis for  $\mf{S}$ exists for any interactions within $\mf{E}$, yet $\mf{S}$ may only interact with all fragments of $\mf{E}$ identically, cf. Fig.~\ref{fig:schemas}. The second part of the analysis is then focused on the dynamics induced by such Hamiltonians that support a pointer basis. We find that the efficiency of the information transfer between $\mf{S}$ and $\mf{E}$ is governed by the statistics of the interaction terms. The average information transfer is irreversible if and only if the support of the coupling coefficients is continuous, and the ``speed of communication'' is determined by the shape of the distribution of the interaction coefficients. Our general findings are illustrated with four models that correspond to a variety of situations, including scrambling or no scrambling, pointer basis or no pointer basis, quantum Darwinism or no quantum Darwinism.

\section{\label{sec:level1}Structure of the Hamiltonian}

We start by defining the problem in mathematically rigorous terms. Consider a set of $(N+1)$ qubits, where the $0$th qubit is the system $\mf{S}$. Hence, the environment $\mf{E}$ is comprised of $N$ qubits. For the sake of simplicity, we further restrict ourselves to 2-body interaction models. The most general Hamiltonian corresponding to this scenario then reads  
\begin{equation}
\label{eq:genH}
H = \sum_{i,j}^{}\sum_{\alpha ,\beta}^{} J^{\alpha \beta}_{ij} \sigma ^{\alpha} _i \otimes  \sigma ^{\beta} _j + \sum_{i}^{}\vec{B_i}\cdot \vec{\sigma _i} ,
\end{equation}
where $\alpha$ and $\beta$ take the values $x$, $y$ and $z$, corresponding to the Pauli matrices $\sigma ^{x}$, $\sigma ^{y}$ and $\sigma ^{z}$. Indices $i$ and $j$ count the qubits, with $i,j=0$ for $\mf{S}$ and $i,j\geq1$ for $\mf{E}$. Further, $J_{ij}$ and $\vec{B_i}$ are real coefficients, which in the following we will choose to be random variables.

\subsection{Existence of a pointer basis} 

The natural question now is, what conditions $J_{ij}$ and $\vec{B_i}$ have to fulfill, such that $H$ in Eq.~\eqref{eq:genH} supports a pointer basis for $\mf{S}$. Pointer states are \emph{the} particular states of $\mf{S}$ that are stable under the interaction with $\mf{E}$ \cite{pointerstates1,pointerstates2,pointerstates3}. Formally, these states can be identified in the following way: $\ket{\psi _\mf{S}} \in \mf{S}$ is a pointer state of $\mf{S}$ if and only if for any $\ket{\psi _\mf{E}} \in \mf{E}$, an initial product state $\ket{\psi _\mf{S}} \otimes \ket{\psi _\mf{E}}$ evolves under $H$ \eqref{eq:genH} to remain within an epsilon ball around the product state $\ket{\psi _\mf{S}}\otimes\ket{\psi _\mf{E}(t)}$. In other words, the reduced state $\ket{\psi _\mf{S}}$ remains pure under the evolution of the total Hamiltonian. 

It will prove convenient to separate the total Hamiltonian into terms corresponding to $\mf{S}$, $\mf{E}$, and their interaction. Hence, we write
\begin{equation}
H = H_\mf{S} \otimes \id_\mf{E}+\id_\mf{S}\otimes H_\mf{E} + H_\mf{SE}\,.
\end{equation}
Comparing with Eq.~\eqref{eq:genH} we identify the system Hamiltonian as
\begin{equation}
\label{eq:sysH}
H_\mf{S}= \vec{B_0}\cdot \vec{\sigma _0}\,,
\end{equation}
whereas we have for the environment
\begin{equation}
\label{eq:envH}
H_\mf{E}= \sum_{1\leq i< j\leq N}^{}\sum_{\alpha,\beta}^{}J^{\alpha \beta}_{ij} \sigma ^{\alpha} _i \otimes  \sigma ^{\beta} _j + \sum_{i=1}^{N}\vec{B_i}\cdot \vec{\sigma _i}\,.
\end{equation}
Notice that the first term in Eq.~\eqref{eq:envH} describes the intra-environmental interactions. The interaction between $\mf{S}$ and $\mf{E}$ is given by
\begin{equation}
\label{eq:intH}
H_\mf{SE} = \sum_{\alpha}^{}\sigma ^{\alpha} _0 \otimes \sum_{j=1}^{N}\sum_{\beta}^{}J^{\alpha \beta}_{0j} \sigma ^{\beta} _j\,.
\end{equation}
From this separation of terms it becomes immediately obvious that the pointer basis for $\mf{S}$ can only exist if certain necessary and sufficient conditions for the interaction term $H_\mf{SE}$ are fulfilled.

These conditions become particularly intuitive by considering the original motivation for \emph{pointer} states. These states are not only immune to the dynamics induced by the interaction with the environment, but can be also thought of as states that correspond to the \emph{pointer} of a measurement apparatus. Mathematically, such an apparatus is described by the pointer observable
\begin{equation} 
A \equiv A_\mf{S}\otimes \mathbb{I}_\mf{E}\,.
\end{equation}
By construction, the pointer observable $A$ commutes with the total Hamiltonian \eqref{eq:genH}, and hence $A$ and $H$ share an eigenbasis. Due to form of $A$ the corresponding eigenstates can be written in tensor-product form  $\ket{S_i} \otimes \ket{E_j}$ with $\left| S_i  \right\rangle \in \mf{S}$ and $\ket{E_j}\in \mf{E}$.

Correspondingly, we can factorize the time-evolution operator as 
\begin{equation}
U = \sum_{i}\ketbrad{S_i}\otimes  \e{-i/\hbar\, H_i\,t}
\end{equation}
where the $H_i$ act only on $\mf{E}$. Now, choosing any (reduced) eigenstate of $A$ as initial state of $\mf{S}$, $\ket{\psi _\mf{S}} = \ket{S_i}$, the product state $\ket{\psi_\mf{S}} \otimes \ket{\psi_\mf{E}}$ evolves by remaining in product form $\ket{\psi _\mf{S}} \otimes \ket{\psi _\mf{E}(t)}$.

In Appendix~\ref{sec:appA} we show that by enforcing the commutation relation $\com{A}{H}=0$ we have that any Hamiltonian \eqref{eq:envH} supporting a pointer basis for $\mf{S}$ has to be of the form
\begin{equation}
H =  H_\mf{S}\otimes \id_\mf{E}+\id_\mf{S}\otimes H_\mf{E} + H_\mf{S} \otimes \sum_{i=1}^{N}h_i ,
\label{structurep}
\end{equation}
where $H_\mf{S}$ is the system Hamiltonian \eqref{eq:sysH}, and the $h_i$ are arbitrary traceless Hermitian operators acting on the $i$th qubit of $\mf{E}$.

In conclusion, we have shown that any model of interacting qubits that supports a pointer basis for a system qubit $\mf{S}$ may have at most a separable interaction term $H_\mf{SE}$. Moreover, this interaction terms has to be factorizable into the system Hamiltonian $H_\mf{S}$ and traceless terms acting on the environmental qubits $\mf{E}$. It is important to emphasize that no additional conditions are required pertaining to, for instance, the intra-environmental interactions. Schematically, our findings are illustrated in Fig.~\ref{fig:schemas}.

\subsection{Further conditions for quantum Darwinism}

It was shown in Ref.~\cite{touil2022branching} that only a special structure of states is compatible with the emergence of quantum Darwinism. These states are of the \emph{singly-branching form} \cite{blume_2005,blume_2006}, which are the only states to support epsilon quantum correlations as measured by quantum discord~\cite{Ollivier2001PRL}.

Singly branching states are pointer states of $\mf{S}$ correlated with the environment states in the special form, 
\begin{equation}
\label{eq:sbs}
\ket{\psi (t)}  = \alpha _0 \ket{0} \bigotimes_{i \in \mf{E}} \ket{\mathfrak{0} _i (t)} + \beta _0 \ket{1} \bigotimes_{i \in \mf{E}}\ket{\mathfrak{1} _i (t)}\,.
\end{equation}
It is easy to see that such a singly branching form can emerge if and only if there are no intra-environmental interactions. 

Thus, we conclude that quantum Darwinism can only be supported by Hamiltonians with separable interaction between $\mf{S}$ and $\mf{E}$, cf. Eq.~\eqref{structurep}, and no intra-environmental interactions, i.e., $J^{\alpha \beta}_{ij}=0$ in Eq.~\eqref{eq:envH}. The remaining question now is whether all such Hamiltonians provide so-called good decoherence, which makes their corresponding $\mf{E}$ good channels for information transfer.

\section{Coefficients of the Hamiltonian}

To analyze the \emph{dynamical} emergence of quantum Darwinism, we now solve for the average dynamics under an arbitrary, random Hamiltonian for which the system $\mf{S}$ has a pointer basis of the single branching form \eqref{eq:sbs}. We will find that the efficiency of information transfer within $\mf{E}$ is governed by the randomness of the interaction coefficients.

\subsection{\label{sec:level2}Solving the dynamics}

To this end, consider an arbitrary Hamiltonian of the form \eqref{structurep}, where we further enforce vanishing intra-environmental interactions $J^{\alpha \beta}_{ij}=0$. Note that in Eq.~\eqref{structurep} the $h_i$ are hermitian, and traceless. Hence, we can write equivalently (and without loss of generality)
\begin{equation}
\label{eq:Hsimple}
H = \sigma^{z} _0 \otimes \sum_{i=1}^{N} B_i\, \sigma ^{z} _i ,
\end{equation}
where the $B_i$ are real random variables.

We are now interested in the dynamics induced by Eq.~\eqref{eq:Hsimple}, and we choose an arbitrary separable initial condition. Therefore, we write
\begin{equation}
\ket{\psi _0} = (\alpha _0 \ket{0} + \beta _0 \ket{1})\, \bigotimes_{i=1}^{N} (\alpha _i  \ket{0_i}+ \beta _i   \ket{1_i}) ,
\end{equation}
where the $\alpha _i$ and $\beta _i$ are arbitrary, complex coefficients. Evolving this $\ket{\psi _0}$ under the corresponding Schr\"odinger equation, $i\partial _t \ket{\psi} = H \ket{\psi}$ we obtain the time-dependent solution,
\begin{equation}
\ket{\psi (t)} = \alpha _0 \left | 0 \right \rangle \bigotimes_{i \in \mf{E}}^{} \left | \mathfrak{0} _i (t)\right \rangle + \beta _0 \left | 1 \right \rangle \bigotimes_{i \in \mf{E}}^{} \left | \mathfrak{1} _i (t)\right \rangle ,
\end{equation}
where we introduced
\begin{equation}
\begin{split}
\ket{\mathfrak{0} _i (t)} &= \alpha _i\, \e{iB_it} \ket{0} + \beta _i\, \e{-iB_it} \ket{1} \\
\ket{\mathfrak{1} _i (t)} &= \alpha _i\, \e{-iB_it}\ket{0} + \beta _i\, \e{iB_it} \ket{1}\,.
\end{split}
\end{equation}

As usual, the reduced density matrix of $\mf{S}$ is given by tracing out $\mf{E}$, $\rho_\mf{S}(t)=\ptr{\mf{E}}{\ket{\psi (t)}\bra{\psi (t)}}$. The corresponding decoherence factor \cite{Riedel_2012} is given by the amplitude of the off-diagonal coefficients of the reduced density matrix $\rho _\mf{S}$ in the basis $\left\{ \ket{0} , \ket{1}\right\}$. We have
\begin{equation}
\la \left | \braket{\mathfrak{1} _i (t) }{\mathfrak{0} _i (t)}\right |^{2} \ra \equiv \la \left | \Gamma _i(t) \right |^{2} \ra\,,
\label{overlap}
\end{equation}
and, since the $B_i$ are stochastically independent, we can write
\begin{equation}
\la \left | \prod\Gamma _i(t) \right |^{2}  \ra = \prod \la \left| \Gamma _i(t) \right |^{2} \ra .
\label{PRODoverlap}
\end{equation}
It is easy to see that we have,
\begin{equation}
\Gamma _i(t) = \left| \alpha _i \right|^{2}\, \e{-2iB_it} + \left| \beta_i \right|^{2}\,\e{2iB_it}\,.
\end{equation}
For random $B_i$ it is, however, more instructive to compute the decoherence factors averaged over all possible values for $B_i$. Further denoting the probability density function of $B_i$ as $P(B_i = x) = f_i(x)$,  we show in Appendix \ref{sec:average} that we obtain
\begin{equation}
\begin{split}
\left \langle \left | \Gamma _i(t) \right |^{2} \right \rangle &= \left| \alpha _i \right|^{4} + \left| \beta_i \right|^{4}\\ &+ 2 \left| \alpha _i \right|^{2} \left| \beta _i \right|^{2} \left| \widetilde{f_i}(4t) \right| \cos(\arg(\widetilde{f_i}(4t))) ,
\end{split}
\label{decoherence_fac_avg}
\end{equation}
where $\widetilde{f_i}$ is the characteristic function of $B_i$
\begin{equation}
\widetilde{f_i}(k) = \int_{-\infty }^{+\infty }dx\,f_i(x)\,\e{ikx}\,.
\end{equation}

In conclusion, we have derived an analytic expression for the average decoherence function, which governs the rate with which information about $\mf{S}$ is communicated through $\mf{E}$.

\subsection{Rate and irreversibility of information transfer}

Equation~\eqref{decoherence_fac_avg} demonstrates the relationship between the emergence of classicality and the randomness of system-environment interactions. Indeed, the probability distributions $f_i$ of the couplings $B_i$ between $\mf{S}$ and $\mf{E}$ play a central role in the rate of information transfer. Observe that the decoherence factors decrease rapidly if $\widetilde{f_i}$ decrease rapidly. Since $\widetilde{f_i}$ is the Fourier transform of the probability distribution $f_i$, the order of differentiability of $f_i$ gives us the order of decay of $\widetilde{f_i}$, while the smallest characteristic length in the distribution $f_i$ gives us the inverse of the characteristic time of decay of $\widetilde{f_i}$.

Furthermore, if the support of $B_i$ is discrete and finite, then the characteristic function $\widetilde{f_i}(k)$ is a periodic (or quasi-periodic) function and therefore does not converge to 0. Hence, having continuous support for the $f_i$ is essential for the emergence of truly classical behavior. In fact, if the $f_i$ distribution is continuous, then the information is transferred \emph{irreversibly}. In this case, the $f_i$ are integrable
\begin{equation}
\int_{-\infty }^{+\infty }dx \,\left | f_i(x) \right | = \int_{-\infty}^{+\infty}dx\,f_i(x) = 1 < \infty ,
\end{equation}
and thus, by virtue of the Riemann-Lebesgue Lemma,
\begin{equation}
\label{eq:epsilon}
\left | \widetilde{f_i}(k) \right |\xrightarrow[k\rightarrow \infty]{}0 \quad\text{and thus}\quad \left \langle \left | \Gamma _i(t) \right |^{2} \right \rangle \xrightarrow[t\rightarrow \infty]{} \epsilon _i < 1.
\end{equation}
where $\epsilon _i$ depends on the initial state. 

Finally, we note that a perfect record of the information about $\mf{S}$ in the $i$th qubit  corresponds to $\Gamma _i = 0$. This is typically not the case. However, as the $\la \left | \Gamma _i(t) \right |^{2} \ra$ become strictly less than one, Eq.~\ref{PRODoverlap} shows that $\ket{0} \bigotimes_{i \in \mf{F}}^{}\ket{\mathfrak{0} _i (t)}$ and $\ket{1} \bigotimes_{i \in F}^{}\ket{\mathfrak{1} _i }$ become orthogonal on average for a sufficiently large fragment $\mf{F}$. Thus, a small set of qubits of the environment is enough to obtain an almost complete record of the state of $\mf{S}$.

\subsection{Quantum Darwinism -- the classical plateau}

\begin{figure}
    \centering
       \includegraphics[width=.48\textwidth]{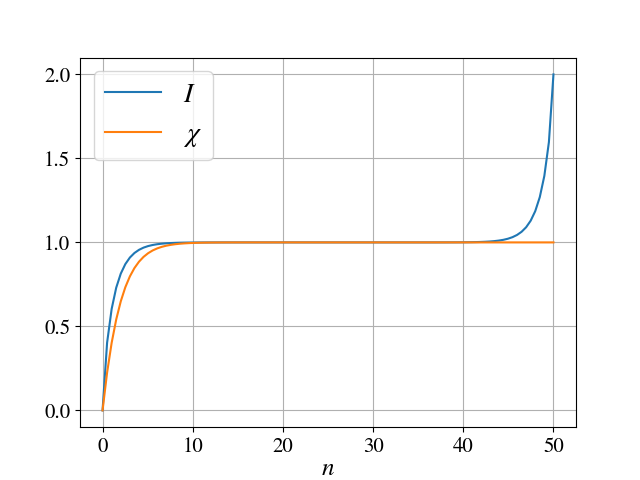}
\caption{\label{fig:plotplateau}Asymptotic mutual information \eqref{eq:I} and Holevo quantity \eqref{eq:chi} as functions of the fragment size. The classical plateau corresponds to the value $\mc{S}_{max} = 1$.}
\end{figure}

The hallmark result of quantum Darwinism is the emergence of the ``classical plateau'' \cite{Zurek2000AP,PhysRevA.72.042113,PhysRevLett.93.220401,RevModPhys.75.715,zurek_2009,girolami_22,touil2022branching,blume_2006,Deffner2023book}, cf. Fig.~\ref{fig:plotplateau}. This classical plateau is a consequence of redundant encoding of the same information in $\mf{E}$.

To this end, consider again a fragment $\mf{F}$ of $\mf{E}$. If any $\mf{F}$ carries the same information about $\mf{S}$, then any two observers accessing different $\mf{F}$ learn exactly the same information about $\mf{S}$. The amount of information that a fragment $\mf{F}$ of $\mf{E}$ contains about the system $\mf{S}$ can be quantified with the mutual information $I(\mf{S}:\mf{F})$ defined as
\begin{equation}
\label{eq:mutI}
I(\mf{S}:\mf{F}) = \mc{S}_\mf{S} + \mc{S}_\mf{F} - \mc{S}_\mf{SF}\,,
\end{equation}
where $\mc{S}(\rho) = -\tr{\rho \log(\rho)}$. The maximal classical information that can be accessed by any observer is upper-bounded by the Holevo quantity \cite{holevo1,holevo2}
\begin{equation}
\chi (\mf{S}:\check{\mf{F}}) = \mc{S}_\mf{S} - \mc{S}_{\mf{S}|\check{\mf{F}}}
\end{equation}
where $\mc{S}_{\mf{S}|\check{\mf{F}}}$ is the conditional von Neumann entropy defined as the minimal von Neumann entropy of $\mf{S}$ obtained after a measurement on $\mf{F}$.

The difference of the mutual information, $I(\mf{S}:\mf{F})$, and the Holevo quantity, $\chi (\mf{S}:\check{\mf{F}})$ has been called quantum discord \cite{Ollivier2001PRL},
\begin{equation}
D(\mf{S}:\check{\mf{F}})\equiv I(\mf{S}:\mf{F})-\chi (\mf{S}:\check{\mf{F}})\geq 0\,.
\end{equation}
Quantum discord measures the genuinely quantum information encoded in $\mf{F}$.

For each $\mf{F}$ and its complement, $\overline{\mf{F}}=\mf{E}\setminus \mf{F}$ we define the corresponding decoherence factors
\begin{equation}
\Gamma _\mf{F} = \prod_{i \in \mf{F}}^{}\Gamma _i \quad\text{and}\quad  \Gamma _{\overline{\mf{F}}} = \prod_{i \notin \mf{F}}^{}\Gamma _i.
\end{equation}
With these definitions, one can then show that for small enough decoherence factors \cite{Riedel_2012} we have
\begin{equation}
I(\mf{S}:\mf{F}) \simeq \mc{S}_{max} - \frac{\xi (\left| \alpha_0 \right|^{2}) }{2} \left[ \left| \Gamma \right|^{2} + \left| \Gamma_{\mf{F}} \right|^{2} - \left| \Gamma_{\overline{\mf{F}}} \right|^{2} \right] ,
\end{equation}
where $\mc{S}_{max} = -\left| \alpha _0 \right|^{2}\log(\left| \alpha _0 \right|^{2}) - (1-\left| \alpha _0 \right|^{2})\log(1-\left| \alpha _0 \right|^{2})$, which is the maximal value that the von Neumann entropy of $\mf{S}$. Correspondingly, we obtain for the Holevo quantity (see Appendix \ref{sec:mut})
\begin{equation}
\chi (\mf{S}:\check{\mf{F}}) \simeq \mc{S}_{max} -  \frac{\xi (\left| \alpha _0 \right|^{2})}{2} \left | \Gamma _\mf{F} \right |^{2}\,,
\end{equation}
which is fully consistent with earlier findings \cite{Touil_2022}.

Finally, in the limit of long times, $t\gg1$, and for smooth enough $f_i$ \eqref{eq:epsilon} we obtain the following asymptotic expression for the mutual information
\begin{equation}
\label{eq:I}
\begin{split}
I(\mf{S}:\mf{F}) \simeq \mc{S}_{max}-  \frac{\xi (\left| \alpha _0 \right|^{2})}{2} \left[ \prod_{i \in \mf{E}}^{}\epsilon _i + \prod_{i \in \mf{F}}^{}\epsilon _i - \prod_{i \notin \mf{F}}^{}\epsilon _i \right]\,,
\end{split}
\end{equation}
and the Holevo quantity
\begin{equation}
\label{eq:chi}
\chi (\mf{S}:\check{\mf{F}}) \simeq \mc{S}_{max} - \frac{ \xi (\left| \alpha _0 \right|^{2})}{2} \prod_{i \in \mf{F}}^{}\epsilon _i\,.
\end{equation}
Further, averaging over every possible separable initial states, we have
\begin{equation}
I(\mf{S}:n)_{\infty} \simeq \mc{S}_{max} -  \frac{\xi (\left| \alpha _0 \right|^{2})}{2}\, \left[ \overline{\epsilon} ^{N} + \overline{\epsilon} ^{n} - \overline{\epsilon} ^{N-n} \right] 
\end{equation}
and
\begin{equation}
\chi (\mf{S}:n)_{\infty} \simeq \mc{S}_{max} -  \frac{\xi (\left| \alpha _0 \right|^{2})}{2}\,\overline{\epsilon} ^{n}\,,
\end{equation}
where $n$ is the size of $\mf{F}$ and $\overline{\epsilon} = 2/3$ (see Appendix \ref{sec:mut}).

These results are depicted in Fig.~\ref{fig:plotplateau} for $N=50$ environmental qubits. Both,  the mutual information and the Holevo quantity exhibit a steep initial rise with increasing fragment size $n$, as larger fragments provide more data about $\mf{S}$.  This initial rise is followed by the classical plateau.

\section{Representative examples}

\begin{table}
\caption{\label{tab:table1}Models classification}
\begin{ruledtabular}
\begin{tabular}{cccc}
 &Pointer Basis&Continuous Support&No Scrambling\\
\hline
CPDI \eqref{eq:CPDI}&\checkmark&\checkmark&\checkmark\\
DPDI \eqref{eq:DPDI}&\checkmark&\textcolor{red}{$\times$}&\checkmark\\
CODI \eqref{eq:CODI}&\textcolor{red}{$\times$}&\checkmark&\checkmark\\
CPDI-S \eqref{eq:CPDI-S}&\checkmark&\checkmark&\textcolor{red}{$\times$}
\end{tabular}
\end{ruledtabular}
\end{table}

We conclude the analysis with the numerical solution of four representative examples.  To support quantum Darwinism a Hamiltonian must obey the following conditions: existence of a pointer basis, continuous support, and no intra-environment interactions. Our first example exhibits these three conditions, and for each following example we successively remove one of these conditions, cf. Tab.~\ref{tab:table1}.

\begin{figure*}

    \begin{subfigure}[b]{1.4in}
      \includegraphics[width=1.4in]{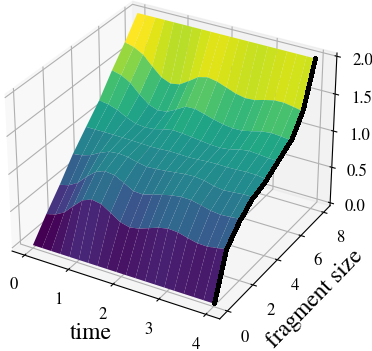}
      \subcaption{}
      \label{fig:plot1}
    \end{subfigure}
    \hspace{0.32cm}
    \begin{subfigure}[b]{1.4in}
      \includegraphics[width=1.4in]{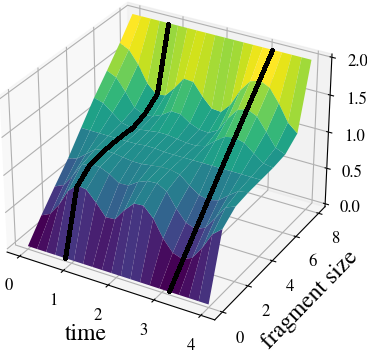}
      \subcaption{}
      \label{fig:plot2}
    \end{subfigure}
    \hspace{0.32cm}
    \begin{subfigure}[b]{1.4in}
      \includegraphics[width=1.4in]{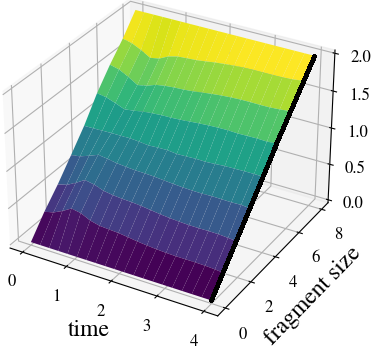}
      \subcaption{}
      \label{fig:plot3}
    \end{subfigure}
    \hspace{0.32cm}
    \begin{subfigure}[b]{1.4in}
      \includegraphics[width=1.4in]{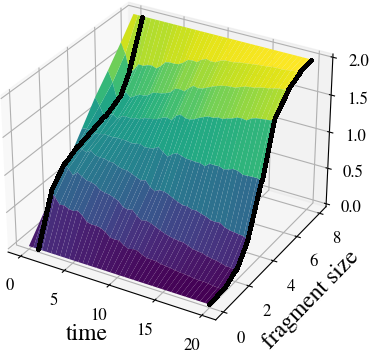}
      \subcaption{}
      \label{fig:plot4}
    \end{subfigure}                     
\caption{Mutual information $I(\mf{S}:\mf{F})$ as a function of time and fragment size for an arbitrary separable initial state of 9 qubits ($N=8$), divided by the von Neumann entropy of $\mf{S}$ (averaged over $10^{2}$ realizations). \textbf{(a)} CPDI \eqref{eq:CPDI}: irreversible information transfer and classical objectivity. \textbf{(b)} DPDI \eqref{eq:DPDI}: non-irreversible (periodic) information transfer. \textbf{(c)} CODI \eqref{eq:CODI}: no local information transfer or redundancy. \textbf{(d)} CPDI-S \eqref{eq:CPDI-S}: competition of emergence of classical objectivity and scrambling in $\mf{E}$.}
\label{fig:plots}
\end{figure*}

\subsection{Continuous Parallel Decoherent Interaction}

The first model has a pointer basis, random coupling coefficients with a continuous spectrum, and does not exhibit scrambling in $\mf{E}$. The corresponding Hamiltonian reads
\begin{equation}
\label{eq:CPDI}
H_\mrm{CPDI} = \sigma ^{z} _0 \otimes \sum_{i=1}^{N}B_i \sigma ^{z} _i
\end{equation}
where $B_i$ are independent random variables,  drawn uniformly from $B_i\in\left [ -1,1 \right ]$. For specificity, we call this model \emph{Continuous Parallel Decoherent Interaction (CPDI)}.

In Fig.~\ref{fig:plot1} we plot the resulting mutual information \eqref{eq:mutI}, as a function of the fragment size, which rapidly converges towards the asymptotic expression \eqref{eq:I}. Note the distinct classical plateau indicative of quantum Darwinism. Moreover, we observe relaxation of $\mf{S}$ into its stationary pointer states over a typical time $\tau \simeq 1/4$, at which point the information transfer becomes irreversible.

\subsection{Discrete Parallel Decoherent Interaction}

Our second example is called, \emph{Discrete Parallel Decoherent Interaction (DPDI)}. The corresponding Hamiltonian is
\begin{equation}
\label{eq:DPDI}
H_\mrm{DPDI} = \sigma ^{z} _0 \otimes \sum_{i=1}^{N}B_i \sigma ^{z} _i
\end{equation}
where $B_i$ are again independent random variables. However, in contrast to the continuous case in Eq.~\eqref{eq:CPDI}, the $B_i$ are now drawn uniformly from the \emph{discrete} set $B_i\in\left \{ -1,-0.5,0.5,1 \right \}$.

In Fig.~\ref{fig:plot2} we depict the resulting mutual information. As expected, we observe that the classical plateau appears and disappears periodically, and hence the information transfer is no longer irreversible. At instants at which the plateau completely vanishes, $I(\mf{S}:\mf{F})$ is linear in the fragment size. This indicates that the information about $\mf{S}$ encoded in $\mf{E}$ is distributed throughout the entire environment (no redunancy).

\subsection{Continuous Orthogonal Decoherent Interaction}

As a next example, we consider \emph{Continuous Orthogonal Decoherent Interaction (CODI)}, which refers to our first model \eqref{eq:CPDI} but with an added external field. This additional term is chosen to be not parallel to the interaction between $\mf{S}$ and $\mf{E}$, and hence CODI does not support a pointer basis. The Hamlitonian reads
\begin{equation}
\label{eq:CODI}
H_\mrm{CODI} = \sigma ^{y} _0 \otimes \mathbb{I}_E + \sigma ^{z} _0 \otimes \sum_{i=1}^{N}B_i \sigma ^{z} _i.
\end{equation}
where, as before, $B_i$ are independent random variables, drawn uniformly from $B_i\in\left [ -1,1 \right ]$.

In this model, information about the observable $\sigma ^{z} _0$ can be registered in $\mf{E}$, but the eigenstates of this observable are not stable and hence not classically objective. This observation is further supported by Fig.~\ref{fig:plot3}, which does not exhibit any form of classical plateau. Moreover, at all instants $I(\mf{S}:\mf{F})$ is a linear function of the size of $\mf{F}$, which is a consequence of the complete absence of any redundancy.

\subsection{Continuous Parallel Decoherent Interaction with Scrambling}

As a final example, we again consider Eq.~\eqref{eq:CPDI} but now design $\mf{E}$ to exhibit scrambling dynamics. Accordingly, this model is called Continuous Parallel Decoherent Interaction with Scrambling (CPDI-S), and the Hamiltonian becomes
\begin{equation}
\label{eq:CPDI-S}
H_\mrm{CPDI-S} = \sigma ^{z} _0 \otimes \sum_{i=1}^{N}B_i \sigma ^{z} _i + \sum_{1\leq i< j\leq N}^{}J_{ij}\sigma ^{z} _i \otimes \sigma ^{z} _j
\end{equation}
where again $B_i$ are independent random variables,  drawn uniformly from $B_i\in\left [ -1,1 \right ]$, and $J_{ij}$ are independent random variables,  drawn uniformly from $J_{ij}\in\left [ -0.03,0.03 \right ]$.

As Fig.~\ref{fig:plot4} shows, a classical plateau rapidly emerges over a time scale of $\tau _\mf{SE} \simeq 1/4$. However, this plateau quickly ``disperses'' as the quantum information becomes non-local due to scrambling in $\mf{E}$. We refer to Ref.~\cite{Riedel_2012} for a more detailed analysis of this particular model. Furthermore, it is interesting to note that Eq.~\eqref{eq:CPDI-S} is another example that demonstrates the competition of decoherence and scrambling as a ``sink for quantum information'' as analyzed by (some of) us in Ref.~\cite{Touil2021PRXQ}.

\section{Concluding remarks}

In the present work we determined the set of qubit models, which support the emergence of classicality. In particular, we established a classification of 2-body interaction models based on the structure of the Hamiltonian and on the nature of its coefficients.

The existence of an ``exact'' pointer basis for the qubit $\mf{S}$ requires the interaction Hamiltonian to be separable between $\mf{S}$ and its environment $\mf{E}$ such that the part acting on $\mf{S}$ is proportionnal to the self Hamiltonian of $\mf{S}$. We call that type of structure a \emph{parallel decoherent interaction}. Furthermore, without any intra-environment interactions, this Hamiltonian structure leads to a branching state structure, the only one compatible with quantum Darwinism \cite{touil2022branching}.

Furthermore, intra-environment interactions can lead to information scrambling in $\mf{E}$, which deteriorates the branching structure. In such situations, the state of $\mf{S}$ is still a classical mixture, but this classicality is hidden to an outside observer who must take a measurement on a non-local part of $\mf{E}$ in order to recover almost all the information about $\mf{S}$. This indicates a clear competition  of the emergence of classical objectivity and scrambling dynamics.

The conceptual notions and the gained insight of our work may open the door for further inquiry, such as the study of $k-$body interactions. In particular, there is every reason to believe that this type of interaction leads to a non-local information transfer. Indeed, for such interactions, the information about $\mf{S}$ is directly encoded in $\mf{E}$ by the entanglement of $\mf{S}$ with fragments $\mf{F}$ of size $k$. This results in a lower redundancy of information. However, the analytical analysis of the dynamics is much more involved than the present 2-interaction case, which is why we leave $k-$body interactions for future work.

\begin{acknowledgments}
This work was carried out during a 15 week internship at the University of Maryland, Baltimore County (P.D.).  We gratefully acknowledge several discussions with Joshua Chiel, who has provided us with perceptive comments. A.T. acknowledges support from the Center for Nonlinear Studies and the U.S DOE under the LDRD program at Los Alamos. S.D. acknowledges support from the John Templeton Foundation under Grant No. 62422.
\end{acknowledgments}

\appendix
\section{\label{sec:appA} Hamiltonians with a pointer basis}

In this appendix we provide further details that lead to the Hamiltonian structure \eqref{structurep}. We start by decomposing the interaction term \eqref{eq:intH} into
\begin{equation}
\label{eq:app1}
H_\mf{SE} = \sigma ^{x} _0 \otimes \sum_{i}^{}\vec{J^{x} _i}\cdot \vec{\sigma _i} + \sigma ^{y} _0 \otimes \sum_{i}^{}\vec{J^{y} _i}\cdot \vec{\sigma _i} + \sigma ^{z} _0 \otimes \sum_{i}^{}\vec{J^{z} _i}\cdot \vec{\sigma _i} .
\end{equation}
Further, we can expand for any pointer observable $A = A_\mf{S} \otimes \mathbb{I} _\mf{E}$ the commutator 
\begin{equation}
\begin{split}
\left [ H,A \right ] &= \left [ H_\mf{S},A_\mf{S} \right ] \otimes \mathbb{I} _\mf{E} + \left [ \sigma ^{x} _0,A_\mf{S} \right ] \otimes H^{x} _\mf{SE}\\ 
&\quad + \left [ \sigma ^{y} _0,A_\mf{S} \right ] \otimes H^{y}_\mf{SE} + \left [ \sigma ^{z} _0,A_\mf{S} \right ] \otimes H^{z} _\mf{SE}\,,
\end{split}
\end{equation}
where $H^{x}_\mf{SE}$, $H^{y} _\mf{SE}$, and $H^{z} _\mf{SE}$ are linear combinations of $\sigma ^{x} _i$, $\sigma ^{y} _i$ and $\sigma ^{z} _i$, which are orthonormal to $\mathbb{I}_\mf{E}$. Thus, a vanishing commutator, $\left [ H,A \right ] = 0$ immediately yields
\begin{equation}
\left [ \sigma ^{x} _0,A_\mf{S} \right ] \otimes H^{x} _\mf{SE} + \left [ \sigma ^{y} _0,A_\mf{S} \right ] \otimes H^{y} _\mf{SE} + \left [ \sigma ^{z} _0,A_\mf{S} \right ] \otimes H^{z} _\mf{SE} = 0 
\end{equation}
as well as $\left [ H_\mf{S},A_\mf{S} \right ] \otimes \mathbb{I} _\mf{E} = 0$\,. Therefore, for any non-trivial $A_\mf{S}$ we must have,
\begin{equation}
H^{x} _\mf{SE} \propto H^{y} _\mf{SE} \propto H^{z}_\mf{SE}\,,
\end{equation}
and the Hamiltonian \eqref{eq:app1} becomes
\begin{equation}
H_\mf{SE} = \vec{J_0} \cdot \vec{\sigma _0} \otimes \sum_{i=1}^{N}\vec{J_i} \cdot \vec{\sigma _i}\,.
\end{equation}
Thus, we have $\left [ \vec{B_0} \cdot \vec{\sigma _0},A_S \right ] = 0 $ and $ \left [ \vec{J_0} \cdot \vec{\sigma _0},A_S \right ] = 0$, which is equivalent to
\begin{equation}
\vec{B_0} \wedge  \vec{J_0} = \vec{0}.
\end{equation}
It means that $\vec{J_0}$ and $\vec{B_0}$ are parallel. We finally obtain the desired result Eq.~\eqref{structurep}.

\section{Average decoherence factors \label{sec:average}}

Equation~\eqref{decoherence_fac_avg} can be obtained by direct derivation. Consider
\begin{equation}
\la \left| \Gamma _i(t) \right|^{2} \ra = \left| \alpha _i \right|^{4} + \left| \beta_i \right|^{4} + 2 \left| \alpha _i \right|^{2}\left| \beta_i \right|^{2} \la  \cos(4B_it)\ra
\end{equation}
where the average is given by
\begin{equation}
\la \cos(4B_i t) \ra = \int dx\,f_i(x)\cos(4xt)\,.
\end{equation}
As above, $f_i(x)$ denotes the probability density of the magnetic field $B_i$. Employing its Fourier transform $\widetilde{f_i}(k)$, we write
\begin{equation}
\la \cos(4B_i t) \ra = \int dx\int dk\,\frac{\widetilde{f_i}(k)}{2 \pi}\e{-ikx}\,\cos(4xt) .
\end{equation}
Now using \cite{abramowitz1948}
\begin{equation}
\int dx\,\e{-ikx}\cos(ax) = \pi\, \left(\delta (k+a) + \delta (k-a)\right),
\end{equation}
we have 
\begin{equation}
\la \cos(4B_i t) \ra= \frac{1}{2}\, \left(\widetilde{f_i}(4t)+\widetilde{f_i}(-4t)\right)\,
\end{equation}
and hence
\begin{equation}
\la \cos(4B_i t) \ra = \left| \widetilde{f_i}(4t) \right|\, \cos(\arg(\widetilde{f_i}(4t)))\,.
\end{equation}

\section{Mutual information and asymptotics \label{sec:mut}}

In this final appendix, we summarize the derivation leading to the asymptotic expressions of the mutual information \eqref{eq:I} and the Holevo quantity \eqref{eq:chi}. We start by considering the reduced density operator of fragment $\mf{F}$, which is given by $\rho _\mf{F} = \ptr{\mf{S}\overline{\mf{F}}}{\rho}$, and we have
\begin{equation}
\rho _\mf{F} (t) = \left | \alpha _0 \right |^{2} \left | F_0 (t) \right \rangle \left \langle F_0 (t)  \right | + \left | \beta _0 \right |^{2} \left | F_1 (t) \right \rangle \left \langle F_1 (t)  \right |\,.
\end{equation}
Explicitly, the states $\ket{F_0 (t)}$ and $\ket{F_1 (t)}$ read
\begin{equation}
\begin{split}
    \left | F_0 (t)\right \rangle &= \bigotimes_{j \in \mf{F}} (\alpha _j \,\e{iB_jt}\left | 0 \right \rangle + \beta _j \,\e{-iB_jt} \left | 1 \right \rangle)\\
\left | F_1 (t)\right \rangle&= \bigotimes_{j \in \mf{F}} (\alpha _j\, \e{-iB_jt}\left | 0 \right \rangle + \beta _j \,\e{iB_jt} \left | 1 \right \rangle) .
\end{split}
\end{equation}
The corresponding decoherence factor is simply given by $\braket{F_1(t)}{F_0 (t)} = \Gamma _\mf{F} (t)$. 

Since we are working with qubits, it is then a simple exercise to show that 
\begin{equation}
\mc{S}_\mf{S} = h\left [ \frac{1}{2}(1+\sqrt{1-4\left | \alpha_0 \right |^{2}\left | \beta_0\right |^{2}(1-\left | \Gamma_\mf{S} \right |^{2})}) \right ]
\end{equation}
and 
\begin{equation}
\mc{S}_\mf{F} = h\left [ \frac{1}{2}(1+\sqrt{1-4\left | \alpha_0 \right |^{2}\left | \beta_0\right |^{2}(1-\left | \Gamma_\mf{F} \right |^{2})}) \right ]
\end{equation}
where
\begin{equation}
h\left [ x \right ] = -x\log(x) - (1-x)\log(1-x)\,.
\end{equation}

These expressions can be further simplified, by expanding for small decoherence factors. We have in leading order
\begin{equation}
\mc{S}_\mf{S} \simeq \mc{S}_{max} -  \frac{\xi (\left| \alpha _0 \right|^{2})}{2} \left | \Gamma \right |^{2}
\end{equation}
and 
\begin{equation}
\mc{S}_\mf{F} \simeq \mc{S} _{max} -  \frac{\xi (\left| \alpha _0 \right|^{2})}{2} \left | \Gamma _\mf{F} \right |^{2}\,,
\end{equation}
where we introduced the notation
\begin{equation}
\xi (\left| \alpha _0 \right|^{2}) = \frac{4\left| \alpha _0 \right|^{2}(1-\left| \alpha _0 \right|^{2})\arctanh (1-2\left| \alpha _0 \right|^{2})}{1-2\left| \alpha _0 \right|^{2}}\, .
\end{equation}

Note that we obtain an equivalent expression for the complement $\overline{\mf{F}}$. Thus, the mutual information \eqref{eq:mutI} becomes 
\begin{equation}
I(\mf{S}:\mf{F}) \simeq \mc{S} _{max} - \frac{ \xi (\left| \alpha _0 \right|^{2})}{2} \left[ \left| \Gamma \right|^{2} + \left| \Gamma _\mf{F} \right|^{2} - \left| \Gamma_{\overline{\mf{F}}} \right|^{2} \right] \,.
\end{equation}
Following similar steps as detailed in Ref.~\cite{Touil_2022} the corresponding Holevo quantity can be written as
\begin{equation}
\begin{split}
\chi (\mf{S}:\check{\mf{F}}) &= h\left [ \frac{1}{2}(1+\sqrt{1-4\left | \alpha_0 \right |^{2}\left | \beta_0\right |^{2}(1-\left | \Gamma \right |^{2})}) \right ]\\
&-h\left [ \frac{1}{2}(1+\sqrt{1-4\left | \alpha_0 \right |^{2}\left | \beta_0\right |^{2}(\left | \Gamma _\mf{F} \right |^{2}-\left | \Gamma \right |^{2})}) \right ]   
\end{split}
\end{equation}
which for weak decoherence in leading order simply is
\begin{equation}
\chi (\mf{S}:\check{\mf{F}}) \simeq \mc{S}_{max} - \frac{ \xi (\left| \alpha _0 \right|^{2})}{2} \left | \Gamma _\mf{F}  \right |^{2}\,.
\end{equation}

Finally, we note that for continuous distributions and $t\gg1$ we have (with Eq.~\eqref{eq:epsilon})
\begin{equation}
I(\mf{S}:\mf{F}) \simeq \mc{S}_{max} - \frac{\xi (\left| \alpha _0 \right|^{2}) }{2} \left[ \prod_{i \in \mf{E}}^{}\epsilon _i + \prod_{i \in \mf{F}}^{}\epsilon _i - \prod_{i \notin \mf{F}}^{}\epsilon _i \right]\, ,  
\end{equation}
where the $\epsilon_i$ are given by
\begin{equation}
\epsilon _i = \left | \alpha _i \right |^{4} + \left | \beta_i \right |^{4}\,.
\end{equation}
Averaging $\epsilon_i$ over all possible values $\alpha _i$ and $\beta _i$ we obtain
\begin{equation}
\overline{\epsilon} = \int_{0}^{1}dx\,(x^{2}+(1-x)^{2}) = \frac{2}{3} .
\end{equation}

\bibliography{sample}

\end{document}